\documentclass[aps,twocolumn,showpacs]{revtex4}
\usepackage{dcolumn}
\usepackage{graphicx}
\usepackage{amsmath}
\usepackage{amsfonts}
\usepackage{amssymb}
\usepackage{psfrag}
\usepackage{wrapfig}
\usepackage{subfigure}
\usepackage{makeidx}
\usepackage{bm}
\usepackage{epsf}

\begin{document}

\title{The Rogue Waves with Quintic
Nonlinearity and Nonlinear Dispersion effects in Nonlinear
Optical Fibers}
\author{Li-Chen Zhao}\email{zhaolichen3@nwu.edu.cn}
\author{Chong Liu}
\author{Zhan-Ying Yang}\email{zyyang@nwu.edu.cn}
\address{$^1$Department of Physics, Northwest University, Xi'an
710069, China}
 %%%%%%%%%%%%%%%%%%%%%%%%%%%%%%%%%%%%%%%%%%%%%%%%%
\date{August 16, 2013}
\begin{abstract}
We present exact rational solution for a modified nonlinear
Schr$\ddot{o}$dinger equation that takes into account quintic
nonlinearity and nonlinear dispersion corrections to the cubic
nonlinearity, which could be used to describe rogue wave in
nonlinear fibers. We find the rogue wave with these higher order
effects has identical shape with the well-known one in nonlinear
Schr$\ddot{o}$dinger equation.  However, the quintic nonlinear term
and nonlinear dispersion effect affect the velocity of rogue wave,
and the evolution of its phase.
\end{abstract}
 \maketitle

\emph{ Introduction}----
Rogue wave(RW) phenomena in the ocean, are
reported to have disastrous consequence, such as destroy ships, oil
platform, etc \cite{Broad,Janssen}. To avoid its negative effects,
people need to know its character, mechanism, even find the ways to
control it. Recently, scientists have done lots of studies on them
\cite{Ruban}. It is shown that it possesses many exceptional
properties, such as much higher than surrounding waves, abrupt
appearance and disappear without any trace, etc. Between the
studies, the nonlinear theory have been paid much attention
\cite{N.Akhmediev,Bludov,Akhmediev,Zakharov,Shrira}. It has been
found that the dynamics equations for RW in ocean, Bose-Einstein
condensate, plasmas, and nonlinear optical fibers are identical
fundamentally. Therefore, the studies on RW in nonlinear fibers
would help us to understand the ones in other systems
\cite{Garrett}. It is well known that the nonlinear
Schr$\ddot{o}$dinger equation (NLSE)
\begin{equation}
i \psi_ z+\psi_{ tt}+ 2 |\psi|^2\psi=0,
\end{equation}
which describes the optical pulse propagation in optics fibers when
the pulse width is greater than 100 femtosecond
\cite{Kibler,Erkintalo,Solli}, Where $\psi=\psi(t, z)$ is the slowly
varying amplitude of the pulse envelope, $z$ represents the distance
along the direction of propagation and $t$ is the retarded time. It
is shown that the rational form solution of the NLSE, could be used
to describe RW well \cite{Shrira}. Furthermore, RW has been observed
recently in one-mode nonlinear fibers experimentally
\cite{Kibler,Erkintalo}, which would highly stimulate RW studies in
a lot of nonlinear systems.

However, the dynamics of these nonlinear systems is significantly
more complicated than the one modeled by the simple NLSE. For
example, for femtosecond optical pulse, higher-order terms that take
into account third-order dispersion, self-steepening and other
nonlinear effects have to be added to this equation \cite{Newell}.
Thus, a question arises: do rogue wave solutions exist for these
more complicated equations? Considering third-order dispersion and
delayed nonlinear response effects, RW in Hirota equation have been
studied in \cite{Ankiewicz,Dai}. Distinct from them, we study RW for
the integrable Kundu-Eckhaus (KE) equation as following \cite{Kundu,
Levi,Radhakrishnan,Geng},
\begin{equation}
 i \psi_ z+\alpha
\psi_{ tt}+\gamma |\psi|^2\psi+4 \beta^2 |\psi|^4 \psi-4 i \beta
(|\psi|^2)_t \psi=0,
\end{equation}
where the subscripts represent the partial derivatives,  $\alpha$ is
the group velocity dispersion coefficient, $\gamma$ is the nonlinear
parameter responsible for the self-phase modulation, $\beta^2$ is
the quintic nonlinearity coefficient, the last term is a nonlinear
term which results from the time-retarded induced Raman process. Eq. (2) has been derived in \cite{Kundu, Levi} and
possesses some applications in the nonlinear optics \cite{Kodama},
quantum field theory \cite{Wang} and weakly nonlinear dispersive
matter waves \cite{Johnson}.

 In this letter, we present exact rational solution for the KE model through Darboux transformation.
 It is found that properties of the rational solution are similar to
 RW's. Therefore, it could be used to describe RW in the model as the previous works \cite{Akhmediev,Zakharov}.
 We find that the quintic nonlinear coefficient and nonlinear
dispersion effect affect the velocity of rogue wave, and change the
evolution of its phase, under the integrable condition, which is the
additional requirement on the coefficients to solve it analytically.
Interestingly, the rogue wave with these higher order effects has
identical shape with the well-known one for NLSE.

 \emph{Exact rational solution and rogue waves}----
The Eq.(2) has
been solved to get soliton solution on trivial background through
Darboux transformation(DT) method in \cite{Geng}.
 As done in NLSE, one can get rational solution
on nonzero plane wave background. We perform the DT method  to
derive rational solution from a plane wave seed solution. With
$\alpha=1$, $\gamma=2$, the corresponding Lax-pair is given in
Appendix part. The nontrivial seed solutions, which can be seen as
the background for RW, are derived as follows

\begin{figure}[htb]
\centering
\subfigure[]{\includegraphics[height=35mm,width=40mm]{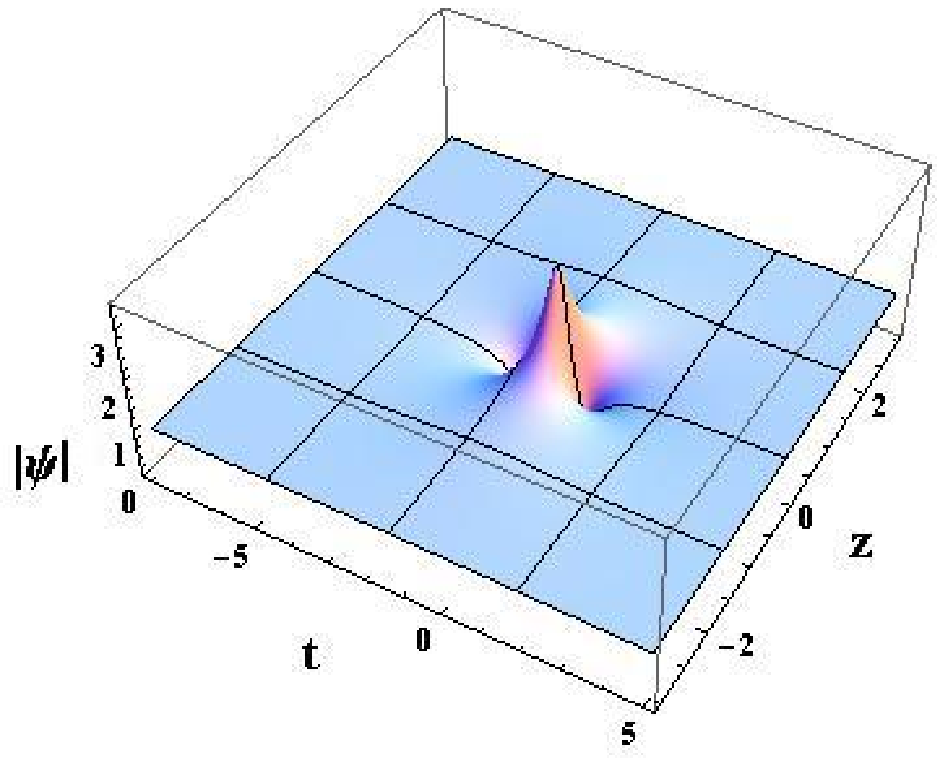}}
\hfil
\subfigure[]{\includegraphics[height=35mm,width=40mm]{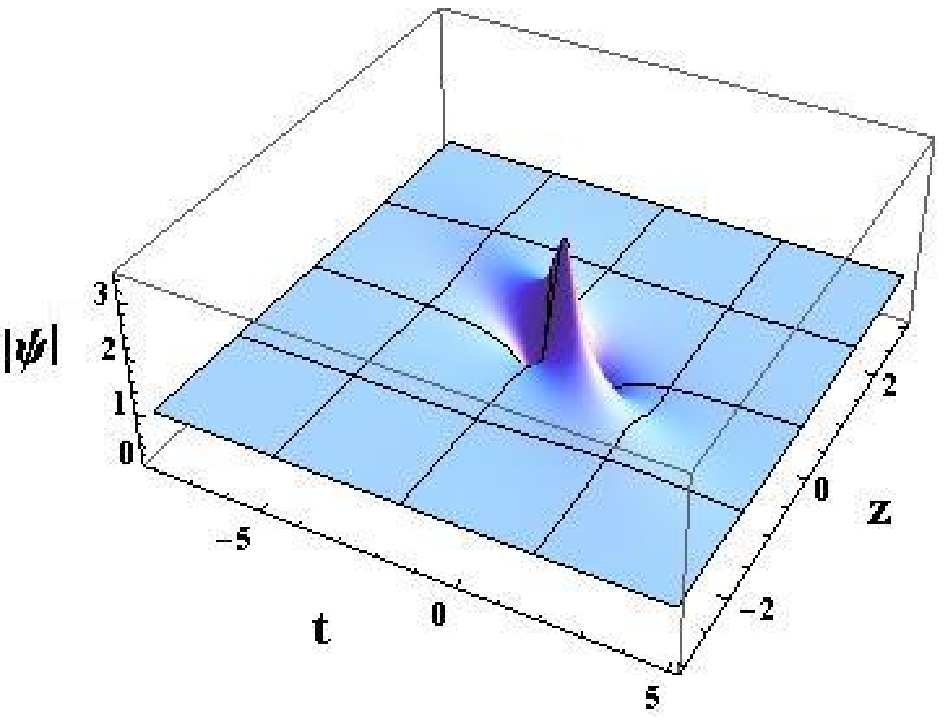}}
\hfil
\subfigure[]{\includegraphics[height=35mm,width=40mm]{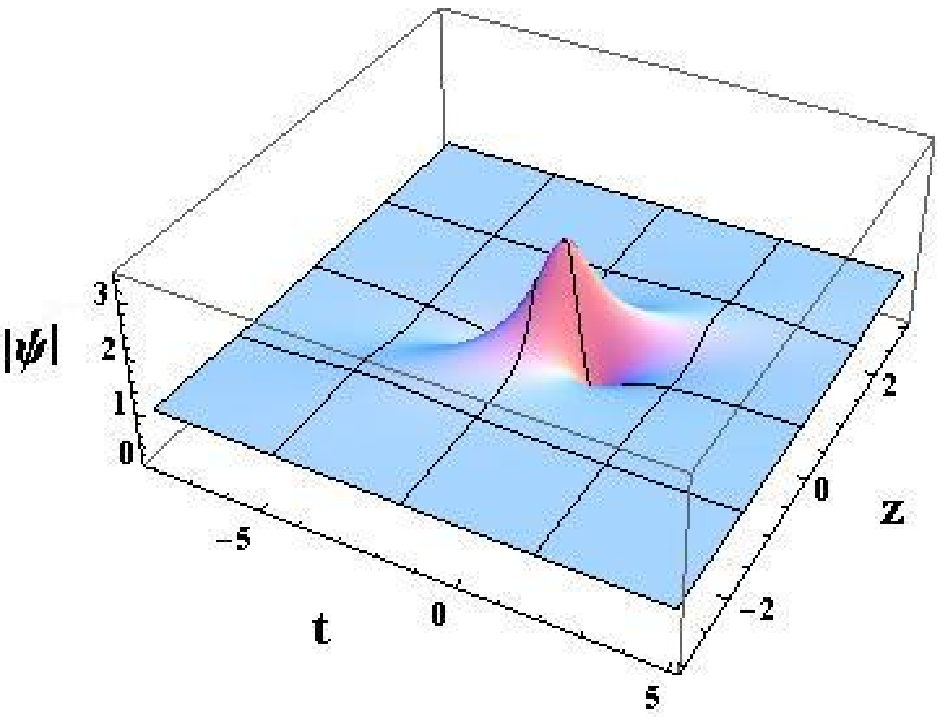}}
\hfil
\subfigure[]{\includegraphics[height=30mm,width=35mm]{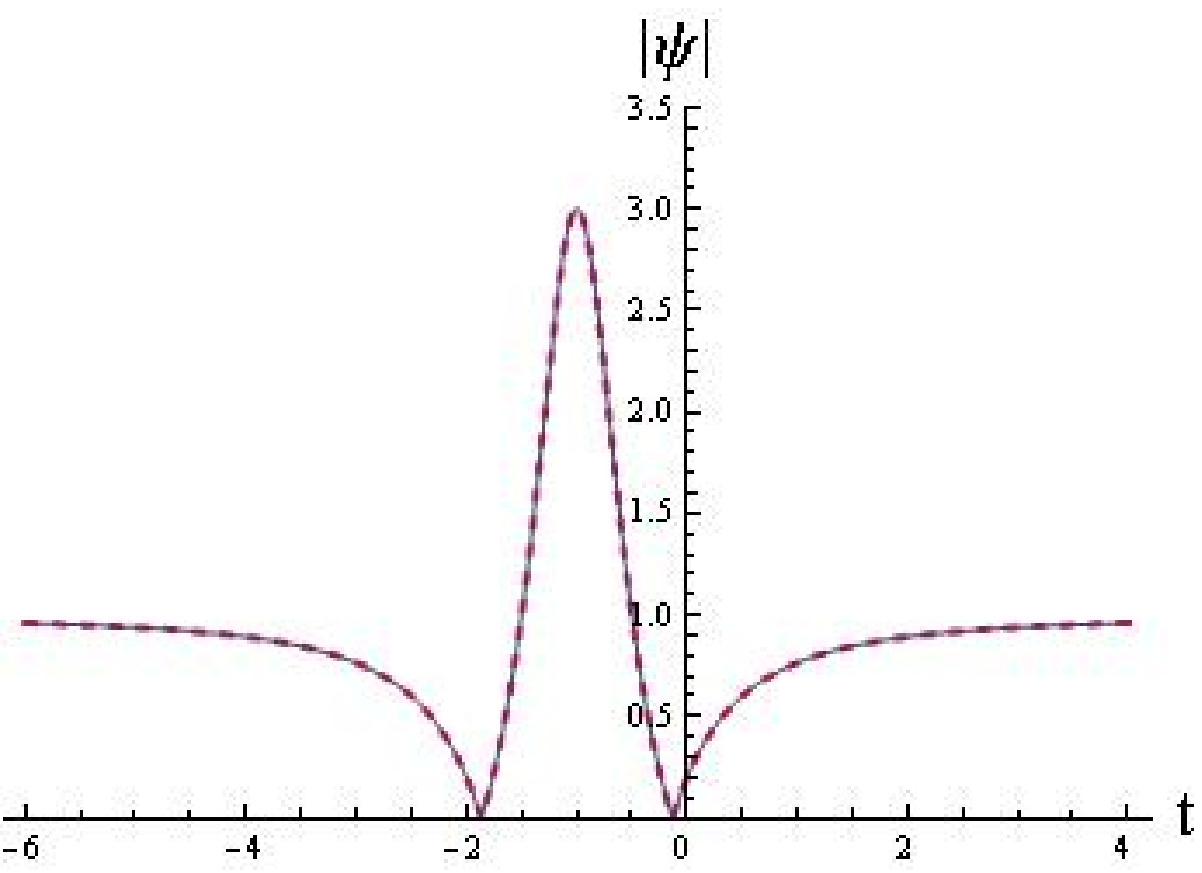}}
\caption{(color online) The evolution of RW with different nonlinear
parameters $\beta$. (a) $\beta=0$; (b) $\beta=0.5$; (c)
$\beta=-0.5$; and (d) the density distribution of RW with different
$\beta$ at $z=0$ (the red dashed one with $\beta=0$, and the blue
one with $\beta=0.5$). It is shown that the density distribution on
time are unchanged with $\beta$, and the parameter $\beta$ just
affect its velocity.}
\end{figure}

\begin{eqnarray}
\psi_{0}&=&s \exp{[i k t+i (4 \beta^2 s^4+2 s^2- k^2)z ]},
\end{eqnarray}
where $s$ and $k$ are two arbitrary real constants, and denote the
amplitude of background and its wave vector respectively. Because
there are trivial transformations connected with different $s$ and
$k$, we can set them $s=1$ and $k=0$ without losing generality. With
spectral parameter $\lambda=\beta+i$ and the nontrivial seed
solutions, we can derive the rational solution as following, through
making the matrix $U$ be Jordan forms,
\begin{eqnarray}
\psi_1=\left[- A^2_1(t,z)+\frac{i 16z+4} {K(t,z)} A^2_1(t,z)\right]
\exp{[i4\beta^2 z+i 2z]},
\end{eqnarray}
where
\begin{eqnarray}
K(t,z)&=&(2t+8\beta z)^2+4(2t+8\beta z)+16z^2+5,\nonumber\\
A_1(t,z)&=&\exp{[4i\beta \frac{2t+8\beta z+2}{(2t+8\beta
z+2)^2+16z^2+1}]}.\nonumber
\end{eqnarray}
 Based on the rational solution, we can study how
quintic nonlinear and nonlinear
  dispersion term affect on RW in nonlinear fibers through varying
  the value of $\beta$. When $\beta=0$, namely, the higher order effects are neglected,
the solution will become the well-known one for NLSE, shown in
Fig.1(a). When $\beta>0$, the RW will have negative velocity on the
retarded time, such as Fig.1(b). When $\beta<0$ it will have
positive velocity, such as Fig.1(c). The velocity of RW increases
with the value of $|\beta|$. Compare the density distribution of the
localized waves, it is found that they have identical shape. As an
example, we show their intensity distribution at $z=0$ in Fig.1(d).
It is pointed that they emerge on different retarded time with the
same shape at $z\neq0$. This indicates that the higher-order effects
affect RW's velocity on retarded time. Furthermore, this character
can be verified through calculating $|\psi_1|^2$ expression.

\begin{figure}[htb]
\centering
\subfigure[]{\includegraphics[height=35mm,width=40mm]{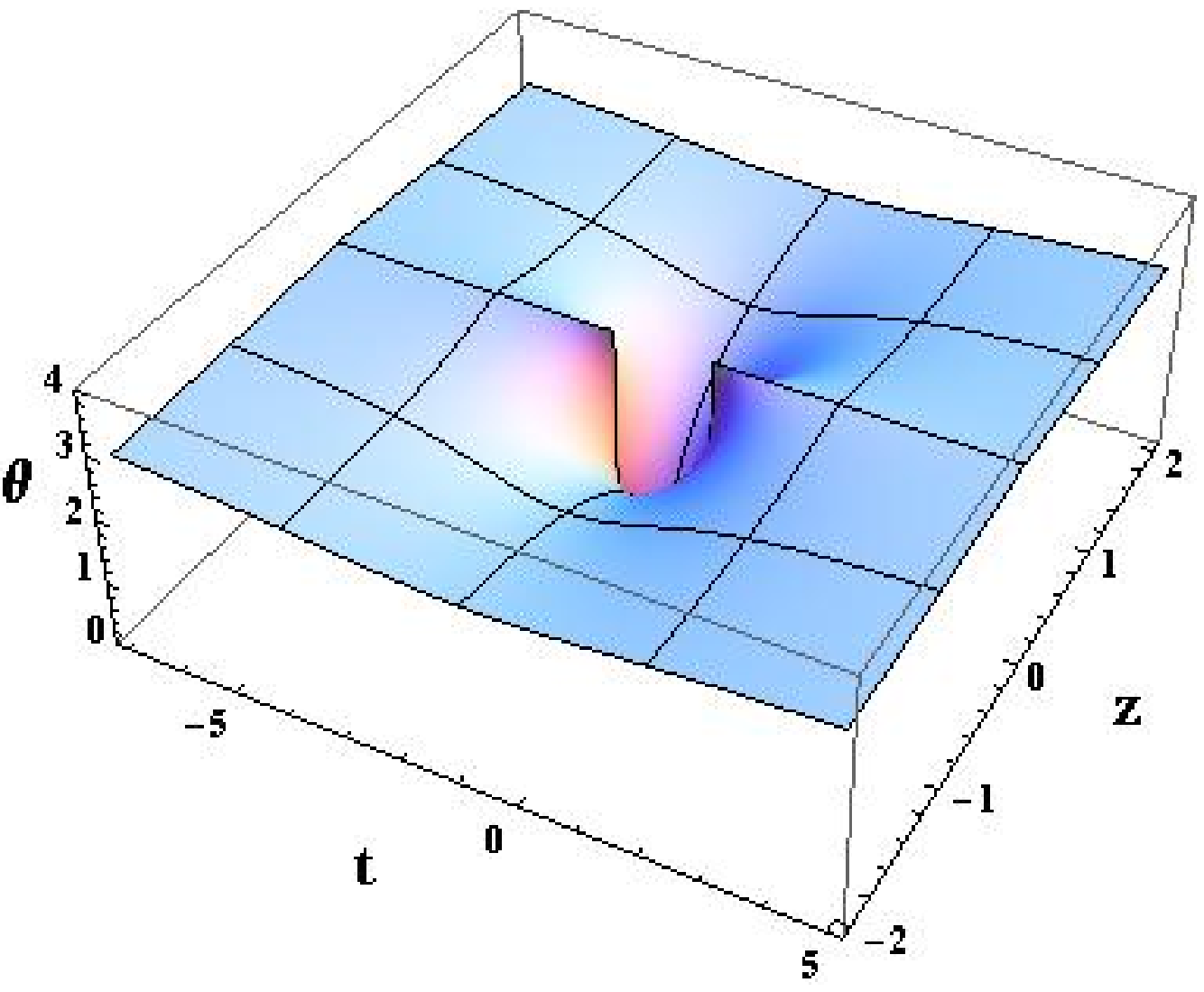}}
\hfil
\subfigure[]{\includegraphics[height=35mm,width=35mm]{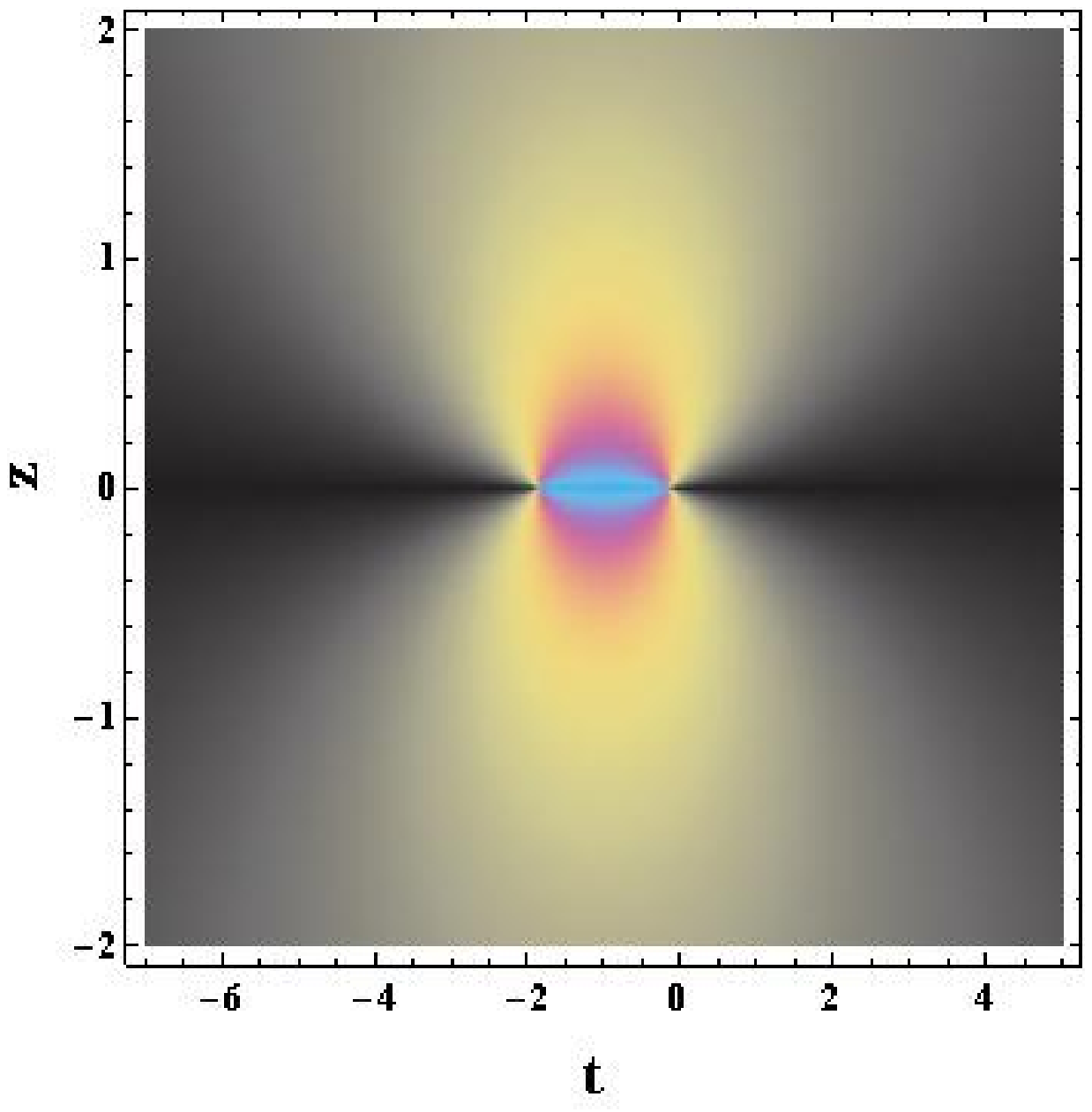}}
\hfil
\subfigure[]{\includegraphics[height=35mm,width=40mm]{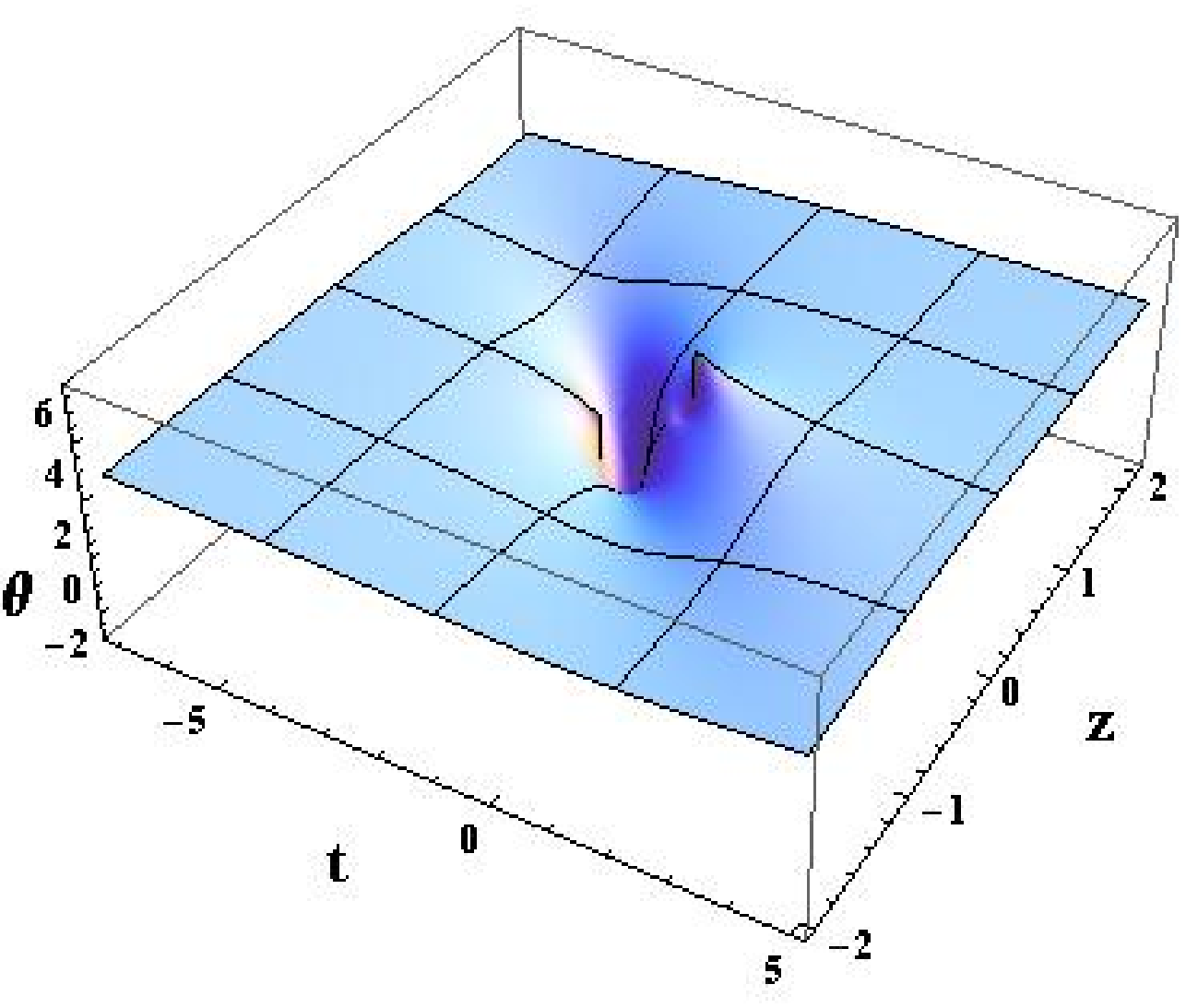}}
\hfil
\subfigure[]{\includegraphics[height=35mm,width=35mm]{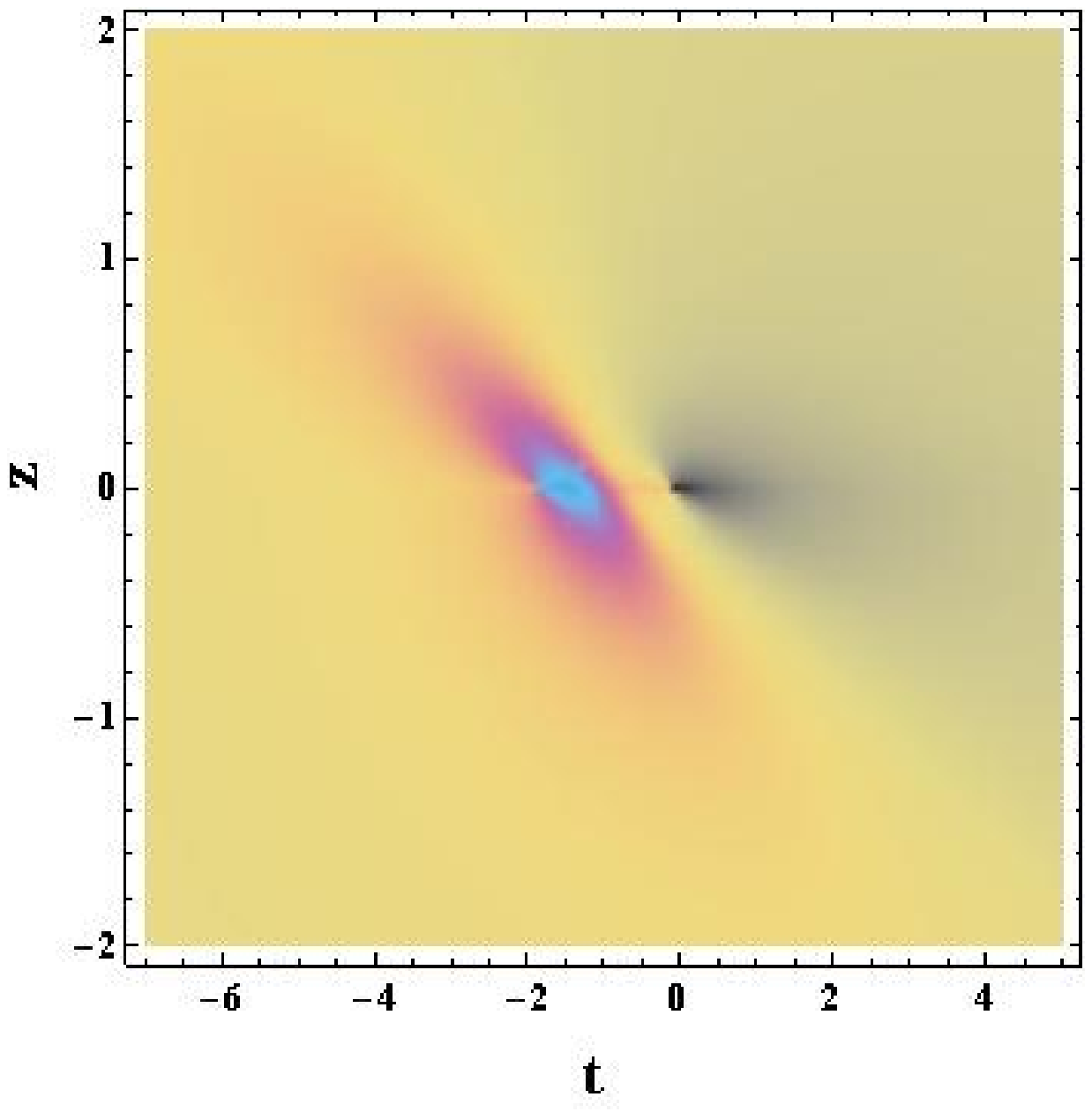}}
\caption{(color online) The evolution of RW's phase with different
nonlinear parameters $\beta$. (a) $\beta=0$, and (b) is the density
plot of (a). (c) $\beta=0.5$, and (d) is the density plot of (c). It
is seen that the symmetric character of RW's is violated with
nonzero $\beta$.}
\end{figure}

Furthermore, we could plot the phase evolution of RW with different
$\beta$. To show the evolution more clear, we ignore the phase of
the background, $(4\beta^2+2)z $, which just increase with
propagation distance $z$. Then, the RW's phase can be given as
\begin{eqnarray}
\theta&=& 8\beta \frac{2t+8\beta z+2}{(2t+8\beta z+2)^2+16z^2+1}\nonumber\\&&+Arccos[\frac{M_1}{\sqrt{M_1^2+M_2^2}}],\\
M_1&=&-1+\frac{4}{K(t,z)},\nonumber\\
M_2&=&\frac{16z}{K(t,z)}.\nonumber
\end{eqnarray}
 The phase evolution of RW with $\beta=0$ is shown in Fig.2(a) and
(b). When $\beta \neq 0$, the phase evolution is shown in Fig.2(c)
and (d). It is seen that the symmetric evolution of RW in NLSE is
broken by the quintic nonlinear coefficient and nonlinear dispersion
effect. Moreover, we find that the phase distribution are opposite
for $\beta$ and $-\beta$ ($\beta \neq 0$).

 \emph{Discussion and Conclusion}----
  In summary, we present exact rational solution for the KE model through Darboux transformation method.
  This indicates that RW can exist with proper higher-order effects.
   Based on
  the analytical solution, it is convenient to study dynamics of RW
  with the cubic and quintic nonlinear terms and nonlinear
  dispersion effects. It is found that RW with these higher-order effects has identical shape with the one of NLSE.
  The quintic nonlinear terms and nonlinear
  dispersion effects just affect the velocity of RW on the retarded time. Moreover, they
  could violate the symmetric evolution of the standard NLSE RW's
  phase. The phase distribution with $-\beta$ is opposite to the one for $\beta$ ($\beta\neq 0$). We believe this character would help us to understand
  properties of RW in many related nonlinear systems with the higher-order effects. However, we
  just investigate this character theoretically, the underlying reason
  is still unknown. It is known that there are always some requirements on the nonlinear
  coefficients to solve it analytically. Maybe these certain
  requirements make related nonlinear effects balance with each
  other. This brings that RW with higher-order term has identical shape
  with standard NLSE one, which just considering second-order
  dispersion term and Kerr nonlinear effect.

 This work has been supported by the National Natural
Science Foundation of China (NSFC)(Grant No.
11047025), and the ministry of education doctoral program
funds (Grant No. 20126101110004).

\section*{Appendix}
The Lax pair of Eq.(2) with $\alpha=1$, $\gamma=2$ could be given as

\begin{eqnarray}
\partial_t\left(
\begin{array}{c}
\Phi_{1} \\ \Phi_{2 }\end{array}
 \right)
&=&U \left(
\begin{array}{c}
\Phi_{1} \\ \Phi_{2}\end{array}
 \right),\\
\partial_z\left(
\begin{array}{c}
\Phi_{1} \\ \Phi_{2}\end{array}
 \right)&=&V\left(
\begin{array}{c}
\Phi_{1} \\ \Phi_{2}\end{array}
 \right),
\end{eqnarray}
where
\begin{equation*}
 U=\left(
\begin{array}{ccc}
-i\lambda+i \beta |\psi|^2&\psi\\
-\bar{\psi}&i\lambda-i \beta |\psi|^2
\end{array}
 \right),
\end{equation*}
\begin{equation*}
 V=\left(
\begin{array}{ccc}
-2i \lambda^2+a_1& 2\lambda \psi+b_1 \\
-2\lambda \bar{\psi}+c_1& 2i \lambda^2-a_1
\end{array}
 \right).
 \end{equation*}
 and
 \begin{eqnarray*}
a_1&=&-\beta (\psi_t \bar{\psi}-\psi \bar{\psi}_t)+4 i\beta^2
 |\psi|^4+i |\psi|^2,\nonumber\\
 b_1&=&i\psi_t+2\beta |\psi|^2 \psi,\nonumber\\
c_1&=&i\bar{\psi}_t-2\beta |\psi|^2 \bar{\psi}.\nonumber
\end{eqnarray*}
 Between the above
expressions,
 the over bar denotes complex
conjugate, and $\lambda$ is spectral parameter.

\end{document}